\documentclass[12pt]{article}
\usepackage{ amsmath, amsthm, amssymb,  }
\usepackage{bbold}
\usepackage{adjustbox}
\usepackage{slashed}

\usepackage{color}
\usepackage{cite}

\usepackage[title,titletoc]{appendix}

\newcommand{\be} { \begin{equation} }
\newcommand{\ee} { \end{equation} }

\newcommand{\eps} {\epsilon}
\begin{document}
\setlength{\unitlength}{1.3cm}
\begin{titlepage}
\vspace*{-1cm}
\begin{flushright}
OUTP-20-14P
\end{flushright}
\vskip 3.5cm
\begin{center}
\boldmath

{\Large\bf Tensor decomposition for bosonic and fermionic scattering amplitudes\\[3mm] }
\unboldmath
\vskip 1.cm
{\large Tiziano Peraro}$^{a,}$
\footnote{{\tt e-mail: tiziano.peraro@unibo.it}} and
{\large Lorenzo Tancredi}$^{b,}$
\footnote{{\tt e-mail: lorenzo.tancredi@physics.ox.ac.uk}}
\vskip .7cm
{\it $^a$ Dipartimento di Fisica e Astronomia, Universit\`a di Bologna e INFN, Sezione di Bologna, via Irnerio 46, I-40126 Bologna, Italy} \\
{\it $^b$
Rudolf Peierls Centre for Theoretical Physics, University of Oxford, Clarendon Laboratory, Parks
Road, Oxford OX1 3PU}
\end{center}
\vskip 2.6cm

\begin{abstract}
In this paper we elaborate on a method to decompose
 multiloop multileg scattering amplitudes into Lorentz-invariant form factors,
which exploits the simplifications that arise from considering four-dimensional external states.
We propose a simple and general approach that applies to both fermionic and bosonic amplitudes
and allows us to identify a minimal number of physically relevant form factors, which
can be related one-to-one to the
independent helicity amplitudes.
We discuss explicitly its applicability to various four- and five-point scattering amplitudes relevant
for LHC physics.

\end{abstract}
\vfill
\end{titlepage}
\newpage


\section{Introduction} \label{sec:intro} \setcounter{equation}{0}
\numberwithin{equation}{section}

In the computation of multiloop multileg scattering amplitudes in quantum field theory, one encounters
at least two different types of  problems, whose complexity increases quickly with the number
of legs and the number of loops involved. First, one has to write down
a loop \emph{integrand}
for the corresponding scattering process in a suitable form and, second, one has to find a way to compute the
relevant \emph{integrals} efficiently, either analytically or numerically.

In this paper we will focus on the first of the two problems.
The standard approach starts from a representation for the scattering
amplitude in terms of Feynman diagrams and requires to perform a series of algebraic manipulations in order
to separate the overall Lorentz structures (i.e.\ combinations of spinors, polarisation vectors, etc\ldots), from
what are properly called \emph{scalar Feynman integrals}. The Feynman integrals
encompass the analytic dependence of the amplitude
on the external kinematics, its branch cuts and its divergences.
While following this program might appear  straightforward at first, it can become
extremely cumbersome in practice for multiloop and multileg scattering amplitudes, where different sets
of manipulations have to be performed on different diagrams in order to achieve the desired decomposition.

A common solution, which allows one to readily decompose any scattering amplitude in terms of scalar
Feynman integrals, is the so-called \emph{projector method}: starting from the symmetries of the scattering amplitudes,
\emph{in primis} Poincar\'e and gauge invariance, one can write down a generic decomposition for any amplitude
in terms of a basis of \emph{Lorentz tensors}, multiplied by scalar form factors.
The decomposition, being only based on symmetry considerations, is non-perturbative and holds at any number
of loops. Starting from this, one can  build projector operators which, once applied on the
scattering amplitude, extract the corresponding form factors.
Very importantly, since loop amplitudes are generically divergent in $d=4$ space-time dimensions and dimensional
regularisation~\cite{Bollini:1972ui,Cicuta:1972jf,tHooft:1972tcz} is used throughout the computation of Feynman integrals,
the aforementioned tensor and projector decomposition is usually  performed in $d$ space-time
dimensions.
After the form factors have been computed, one can use them to compute the so-called
\emph{helicity amplitudes} for the process considered. In practice, one identifies the set of independent
helicity amplitudes necessary to fully characterise the problem and, starting from the generic $d$-dimensional
tensor decomposition, one fixes the helicities of the external states, obtaining a representation for
the helicity amplitudes in terms of linear combinations of the form factors. Clearly,
in doing this, one has to switch from $d$- to four-dimensional external states, moving from
what is usually referred to as Common Dimensional Regularisation (CDR) to the so-called
't Hooft-Veltman scheme (tHV).\footnote{See~\cite{Gnendiger:2017pys} and references therein,
for a discussion about different regularisation schemes.}
In this step one often realises that only specific
linear combinations of the original form factors appear in the
physical helicity amplitudes, often  leading
to simplifications in the final result.

This method has been applied extremely successfully in the calculation of a
countless scattering amplitudes for $2 \to 1$
and $2 \to 2$ processes, see for example~\cite{Garland:2002ak,Glover:2003cm,Glover:2004si,Gehrmann:2015ora}.
Nevertheless, there are (at least) two main reasons why one might want to improve on this method.
First of all, with the increase of the number of external legs, the number of $d$-dimensional tensors required
is bound to increase very fast due to the  combinatorics stemming from the large number of independent
external momenta. Moreover, insisting on using $d$-dimensional external states, when dealing
with external fermions, forces us to work with the $d$-dimensional algebra of gamma matrices
which, differently from the four-dimensional one, is not closed. As we will see in an explicit example below,
for some processes this
prevents us to even be able to provide a generic non-perturbative tensor decomposition
valid at any number of loops.
Finally, it seems  reasonable to expect that if we were able to identify from the
beginning those and only those (combinations of)
form factors that contribute to the physical scattering amplitudes, their calculation could turn out to be
simpler than that of the the full, unphysical, $d$-dimensional ones.

This is not the first paper where this question has been addressed. In Ref.~\cite{Chen:2019wyb}
a solution was proposed, which involves the computation of polarised scattering amplitudes
in an hybrid dimensional regularisation scheme.  In the context of this method, a connection between helicity amplitudes and a four-dimensional treatment of tensors and projectors has also been highlighted, in Ref.~\cite{Ahmed:2019udm}. A different idea
has been proposed  in
Ref.~\cite{Peraro:2019cjj}, where it has been shown that one can build projector operators that
project directly onto the helicity amplitudes. The important finding of \cite{Peraro:2019cjj} was that, when
applying this idea to the scattering of $n > 4$ particles, the ensuing helicity projectors could be expressed
as linear combinations of a physical subset of the standard $d$-dimensional tensors.
Interestingly, the number of independent tensors would match  the number
of  helicity amplitudes, pointing to the fact that the minimal complexity of the problem
could be exposed.
As a consequence, restricting to the subspace of physical four-dimensional tensors allows one
to keep the complexity
stemming from the large number of independent $d$-dimensional tensors under control,
since their number is always bound by the number of helicity amplitudes in the problem.

Two important considerations were still missing in~\cite{Peraro:2019cjj}: first, how to generalise those
findings to (the in principle simpler case of) the scattering of $n \leq 4$ particles; secondly, an efficient
way to determine the basis of independent tensors in the fermionic case, which does not go through the
enumeration of all
$d$-dimensional tensor structures and the explicit construction of helicity projectors.
In this paper we provide a solution for these two remaining issues, describing  a new
generic approach to the decomposition of multiloop multileg Feynman amplitudes in the 't Hooft-Veltman
scheme, which matches one to one the number of independent helicity
amplitudes in the problem.

\section{The general idea}
\label{sec:general} \setcounter{equation}{0}
\numberwithin{equation}{section}

In order to describe the  idea behind this paper, let us consider a generic process, where
$n$ unspecified bosonic and/or fermionic  particles collide.
For definiteness we could imagine to be working in QCD,
but this is not necessary for the discussion that will follow.
Using Lorentz invariance, gauge invariance and any other symmetries of the problem at hand, we can parametrise
the scattering amplitude for the process in terms of a basis of independent tensor structures $T_i$ as
\begin{equation}
\mathcal{A}(p_1,...,p_n) = \sum_{i=1}^N \mathcal{F}_i(p_1,...,p_n)\, T_i\,, \label{eq:amplgen}
\end{equation}
where $\mathcal{F}_i$ are scalar form factors. The tensors $T_i$ are built using the polarisation states of the
external particles, respecting the Lorentz and gauge symmetries of the problem, and assuming that external states
are $d$-dimensional.
The standard procedure would then consist in defining projector operators, decomposed in terms of the same basis tensors
\begin{align}
\mathcal{P}_j = \sum_{i=1}^N c_{i}^{(j)}(d;p_1,...,p_n) T^\dagger_i\,,\label{eq:defPj}
\end{align}
such that by applying them on the amplitude in eq.~\eqref{eq:amplgen} and summing over the polarisations of the external states
one finds
\begin{align}
\sum_{pol}\, \mathcal{P}_j \mathcal{A} (p_1,...,p_n) = \mathcal{F}_j(p_1,...,p_n)\,. \label{eq:sumpolPj}
\end{align}
The coefficients $c_{i}^{(j)}(d;p_1,...,p_n)$ in eq.~\eqref{eq:defPj} can be computed by imposing that eq.~\eqref{eq:sumpolPj} is satisfied.
This requires in general to solve a system of $N$ equations in $N$ unknowns.
Equivalently, it is convenient to define the matrix
\begin{equation}
M_{ij} = \sum_{pol}T_i^\dagger T_j \label{eq:matrixproj}
\end{equation}
such that the projectors can be obtained by computing its inverse
\begin{equation}
 c_{i}^{(j)}(d;p_1,...,p_n) = \left( M^{-1} \right)_{ij}\,. \label{eq:cinM}
\end{equation}
We stress here that in general the solution  for $c_{i}^{(j)}(d;p_1,...,p_n)$ will
depend on the space-time dimensional regulator $d$ and might in particular be singular as $d \to 4$.

Let us now go back to the amplitude in eq.~\eqref{eq:amplgen}. In order to compute physical observables,
we are ultimately interested in considering so-called helicity amplitudes,
which are obtained by fixing the helicities of the external states in all possible ways. This can be done directly
in  eq.~\eqref{eq:amplgen},
where the only dependence on the polarisation of the external states resides in the tensors $T_i$.
In particular,
if the helicity of particle $i$ is $\lambda_i$, we can write
\begin{equation}
\mathcal{A}^{\lambda_1,...,\lambda_n}(p_1,...,p_n) = \sum_{i=1}^N \mathcal{F}_i(p_1,...,p_n)\,
T_i^{\lambda_1,...\lambda_n}\,. \label{eq:helampl}
\end{equation}
Importantly, when we fix the helicities we also specify the external
states to be four-dimensional.
This calculation scheme, where the form factors are computed in dimensional regularisation but the external
states are kept four-dimensional, corresponds to the so-called 't
Hooft-Veltman scheme, where external momenta and polarizations are in
four dimensions while internal states and loop momenta are in $d$ dimensions.
It is not difficult to imagine that when this is done, some of the tensors that used to be independent for generic $d$
can become linearly dependent of each other.
It is easy to see that this should happen whenever the number of $d$-dimensional tensors $N$ is larger than the number
of helicity amplitudes with four-dimensional external states. In this case, in fact, one could imagine to promote the linear combinations of
tensors in eq.~\eqref{eq:helampl} to new \emph{independent tensors} and one would expect that only
those combinations should be enough to describe the helicity amplitudes.
Here it is important to stress that the counting of the independent
helicity amplitudes slightly changes depending on the number
of external legs in the process. Indeed, it is easy to convince
oneself that, with $n \leq 4$ external particles in four dimensions, it is impossible to construct a parity-odd invariant free of spinor phases.
This implies that the helicity amplitudes for such processes, up to an
overall helicity-dependent spinor phase which does not depend on the
loop order, have trivial
behaviour under parity transformation.  This allows one to reduce the number of different helicity amplitudes by a factor two.
On the other hand, for $n \geq 5$ external particles one can build parity-odd invariants and the
helicity amplitudes will depend in a non-trivial way on them. The typical example is the scattering of five massless
partons in QCD where the amplitudes depend on the parity odd invariant
${\rm tr}5 = {\rm tr} ( \gamma_5 \slashed{p}_1\slashed{p}_2\slashed{p}_3\slashed{p}_4 )$.
In this case all helicities have to be treated as different.

We will see various  examples of this in the following. For now, let us ignore the details
and simply assume that, when four-dimensional external states are considered, only the first $Q \leq N$
tensors are independent \emph{and} they are sufficient to span the tensor space on which the scattering amplitude is defined.
As we will see more explicitly later, one way the linear dependence manifests in our calculation is through
the fact that the determinant of the matrix in eq.~\eqref{eq:matrixproj} would go to zero as $d \to 4$.
In order words, in $d = 4$, the matrix $M$ would not be full rank, ie $rank(M) = Q \leq N$.
In this situation, we  can always reorder the tensors such that the independent ones are the first $Q$
\begin{align}
\overline{T}_j = T_j\,, \quad j=1,...,Q \,.
\end{align}
We also assume that the first $Q$ tensors are sufficient to describe the
tensor space spanned by the external helicity states.  It is then convenient to define $Q$ \emph{intermediate projector operators} given by
\begin{equation}
M^{Q\times Q}_{ij} = \sum_{pol}\overline{T}_i^\dagger \overline{T}_j \,, \qquad
P^{Q\times Q}_i =  \sum_{j=1}^Q \left(M^{Q\times Q}\right)^{-1}_{ij} \, \overline{T}_j^\dagger\,, \qquad i,j=1,...,Q\,. \label{eq:matrixprojQ}
\end{equation}
The matrix $M^{Q\times Q}_{ij}$ is computed and inverted in $d$ dimensions, but by construction it
has full rank even in the limit $d \to 4$.

At this point, we  complete the basis of $d$-dimensional tensors by defining the remaining $N-Q$ tensors
\begin{equation}
\overline{T}_i = T_i - \sum_{j=1}^Q \left( P^{Q\times Q}_j\,  T_i \right) \overline{T}_j\,, \qquad i=Q+1,...,N\,, \label{eq:orthogonalTj}
\end{equation}
where, once more, projections and sums over polarisations are performed in $d$ dimensions.
The definition in eq.~\eqref{eq:orthogonalTj} amounts to removing from the original $T_i$  their
projection along the first $Q$ tensors.
Of course, if we are only interested in the $d$-dimensional problem,
it is always possible to perform such a re-definition of the basis of tensors which will
effectively block-diagonalise the corresponding projectors. Indeed, we can now define
\begin{equation}
\overline{M}_{ij} = \sum_{pol}\overline{T}_i^\dagger \overline{T}_j \,, \qquad
\overline{P}_i =  \sum_{j=1}^N \left(\overline{M} \right)^{-1}_{ij} \, \overline{T}_j^\dagger\,,  \qquad i,j=1,...,N\,, \label{eq:matrixprojFinal}
\end{equation}
where the inverse of the matrix $\overline{M}$ will have, by construction, the block form
\begin{equation}
\left(\overline{M}\right)^{-1}_{ij} = \left(
\begin{array}{cc}
\left(M^{Q\times Q}\right)^{-1}_{ij} & 0  \\
0 & Y_{ij}
\end{array}
\right)\,.\label{eq:matrixprojBlock}
\end{equation}

Until now we have performed all manipulations in $d$ dimensions and have
not made explicit use of the fact that the first $Q$ tensors span the entire physical space required to
compute the helicity amplitudes in for four-dimensional external states.
 Let us go now back to eq.~\eqref{eq:amplgen}. It is straightforward to re-express it in terms of the new basis of tensors
\begin{align}
\mathcal{A}(p_1,...,p_n) &= \sum_{i=1}^Q \mathcal{F}_i(p_1,...,p_n)\, T_i + \sum_{i=Q+1}^N \mathcal{F}_i(p_1,...,p_n)  T_i \,, \nonumber \\
&=\sum_{i=1}^Q \left[ \mathcal{F}_i +  \sum_{j=Q+1}^N \mathcal{F}_j \left( P^{Q\times Q}_i\,  T_j \right)\right]\, \overline{T}_i
+ \sum_{j=Q+1}^N \mathcal{F}_j\, \overline{T}_j
\quad = \quad \sum_{i=1}^N \overline{\mathcal{F}}_i\, \overline{T}_i
 \label{eq:amplgenV2}
\end{align}
where
\begin{align}
\overline{\mathcal{F}}_i =
\left\{ \begin{array}{cc}
 \mathcal{F}_i +   \sum_{j=Q+1}^N \mathcal{F}_j \left( P^{Q\times Q}_i\,  T_j \right) \, & i =1,...,Q \\
  \mathcal{F}_i\, &  i =Q+1,...,N\,.
\end{array} \right.
\label{eq:newFs}
\end{align}
From Eq.~\eqref{eq:newFs}
we see that, as long as the matrix $M^{Q\times Q}_{ij}$
defined in eq.\eqref{eq:matrixprojQ}
has full rank in $d=4$ (and therefore its inverse is finite in $d= 4$), the new form factors
$\mathcal{\overline{F}}_i$ are smooth linear combinations of the
original form factors $\mathcal{F}_i$.
We proceed now and fix the helicities on eq.~\eqref{eq:amplgenV2}
finding
\begin{equation}
\mathcal{A}^{\lambda_1,...,\lambda_n}(p_1,...,p_n) = \sum_{i=1}^Q \mathcal{\overline{F}}_i(p_1,...,p_n)\,
\overline{T}_i^{\lambda_1,...\lambda_n} \label{eq:helamplFinal}\,,
\end{equation}
where we stress that the sum runs only over the first $Q$ tensors.
We stress that this is the crucial point: as long as the first $Q$ tensors span the full tensor space
defined by the external states, the $N-Q$ remaining tensors defined in eq.~\eqref{eq:orthogonalTj}
are exactly zero for four-dimensional
external states
and, in particular, they are zero when we fix the helicities of the external particles in all possible ways
\begin{equation}
\overline{T}_i^{\lambda_1,...\lambda_n} = 0\,, \qquad i =Q+1,...,N\,.
\end{equation}
From now one we will refer to these tensors as \emph{irrelevant tensors} and similarly to their projectors as
\emph{irrelevant projectors}, while the non-zero ones will be refereed to as \emph{relevant} ones.


Let us pause here to stress what we have achieved: we have found a new decomposition of the amplitude
which allows us to extract the full information necessary for computing all physical
helicity amplitudes from a minimal number of form factors which multiply tensors that are independent in $d=4$
space-time dimensions. As we will see explicitly below, in our construction it is crucial that
the tensors that we pick span the full tensor-space where the scattering amplitude is defined.
Intuitively, we expect the dimension of this space to coincide with the number of different helicity amplitudes
for the process considered, which implies that there is a one-to-one correspondence between the helicity amplitudes
and our relevant tensor structures.
Notice also that, since the matrix in eq.~\eqref{eq:matrixprojBlock} is in block form,
the $Q$ physical form factors can be computed using projectors that are decomposed purely
in terms of the $Q$ independent tensors:
we never need to compute the contractions of the amplitude with the irrelevant tensors $\overline{T}_{i}$, $i=Q+1,...,N$.
While these properties are clearly appealing from an aesthetic point of view, depending on the complexity of the problem at hand they
 might also constitute a source of substantial practical simplification.

 In the next section we will show how the formal construction works out in practice for different $4$- and $5$-point scattering processes,
 trying to elucidate all points discussed above.

\section{Case of study: $gg \to gg$ scattering}
\label{sec:gggg} \setcounter{equation}{0}
\numberwithin{equation}{section}

We start by considering the scattering of four massless
spin-1 bosons. For definiteness we imagine to deal with
 four-gluon scattering, but this is not necessary for the arguments that follow.
Let us consider the process
\begin{equation}
g(p_1) + g(p_2) + g(p_3) + g(p_4) \to 0\,,\label{eq:4g}
\end{equation}
and call $\epsilon_j^\mu$ the polarisation vector
for to the gluon $j$. We define the usual Mandelstam invariants to be
\begin{equation}
s=(p_1+p_2)^2\,, \qquad t = (p_1+p_3)^2\,, \qquad u = (p_2+p_3)^2 = -s-t\,. \label{eq:mandelstam4}
\end{equation}
We indicate the scattering amplitude as $\mathcal{A}_{4g}(p_1,p_2,p_3)$.
Once stripped of the polarisations of the external gluons, this amplitude must be a rank-4 Lorentz tensor
\begin{equation}
\mathcal{A}_{4g}(p_1,p_2,p_3) = \epsilon_{1,\mu}\epsilon_{2,\nu}\epsilon_{3,\rho}\epsilon_{4,\sigma}\;  \mathcal{A}_{4g}^{\mu \nu \rho \sigma}(p_1,p_2,p_3)\,.
\end{equation}
The most general Lorentz-covariant decomposition for a rank-4 tensor in $d$ space-time dimensions
involves $138$ tensor structures, see for example~\cite{Binoth:2002xg}.
Imposing transversality for each external gluon
$\epsilon_i \cdot p_i = 0$, together with fixing the gauge, allows one to reduce these structure to
$10$ independent ones.
In the case of gluon scattering, given the symmetry of the external states and the fact that color ordering can be used to isolate simpler
primitive amplitudes, a convenient gauge choice is the cyclic one $\epsilon_i \cdot p_{i+1} = 0$, where we identify $p_5 = p_1$.
With this choice the 10 tensors read
\begin{align}
&T_1 = \epsilon_1 \cdot p_3\, \epsilon_2 \cdot p_1\, \epsilon_3 \cdot p_1\, \epsilon_4 \cdot p_2 \,, \nonumber \\
&T_2 = \epsilon_1 \cdot p_3\, \epsilon_2 \cdot p_1\, \epsilon_3 \cdot \epsilon_4 \,, \quad
T_3 = \epsilon_1 \cdot p_3\, \epsilon_3 \cdot p_1\, \epsilon_2 \cdot \epsilon_4 \,, \quad
T_4 = \epsilon_1 \cdot p_3\, \epsilon_4 \cdot p_2\, \epsilon_2 \cdot \epsilon_3 \,, \nonumber \\
&T_5 = \epsilon_2 \cdot p_1\, \epsilon_3 \cdot p_1\, \epsilon_1 \cdot \epsilon_4 \,, \quad
T_6 = \epsilon_2 \cdot p_1\, \epsilon_4 \cdot p_2\, \epsilon_1 \cdot \epsilon_3 \,, \quad
T_7 = \epsilon_3 \cdot p_1\, \epsilon_4 \cdot p_2\, \epsilon_1 \cdot \epsilon_2 \,, \nonumber \\
&T_8 = \epsilon_1 \cdot \epsilon_2\,  \epsilon_3 \cdot \epsilon_4 \,, \quad
T_9 = \epsilon_1 \cdot \epsilon_4\,  \epsilon_2 \cdot \epsilon_3 \,, \quad
T_{10} = \epsilon_1 \cdot \epsilon_3\,  \epsilon_2 \cdot \epsilon_4 \,. \label{eq:tensors4gddim}
\end{align}
Following the notation introduced in section~\ref{sec:general}, we then write the amplitude for the scattering of four gluons as
\begin{equation}
\mathcal{A}_{4g}(p_1,p_2,p_3) = \sum_j \mathcal{F}_j(p_1,p_2,p_3) \; T_j
\end{equation}
where $F_j$ are the 10 scalar form factors.
In CDR (i.e. keeping the external polarisations in $d$-dimensions) these 10 tensors are
independent and they must be kept as they are.
As described in the previous section,
the standard approach would then be to derive 10 projectors that single out each of the form factors.

Nevertheless, we have seen that as long as not all tensors are independent for four-dimensional
external states,
this procedure can be simplified. A first counting of the number of helicities
involved shows immediately that we expect $2^4 = 16$ helicities, which are reduced to $8$ by
parity (we recall, from the previous section, that this reduction by
parity holds up to $n = 4$ external
legs). We expect, therefore, to be able to describe the physical process
in terms of $8$ independent tensor structures.
Indeed, 2 of the tensors in eq.~\eqref{eq:tensors4gddim} do
become linear dependent when the external states are specialised in four dimensions.
This can be demonstrated in several ways. One possibility is to explicitly project the $10$ tensors in eq.~\eqref{eq:tensors4gddim}
onto a complete set of independent external four-dimensional helicity states  (for
example using the spinor helicity formalism).

As hinted to in section~\ref{sec:general}, an alternative way to see this
is to study the rank of the matrix
$$M_{ij} = \sum_{pol}T_i^\dagger T_j$$
where in the case under study the sum over polarisations has to be performed respecting
\begin{equation}
\sum_{pol} \epsilon_i^\mu \epsilon_i^\nu = - g^{\mu \nu} + \frac{p_i^\mu p_{i+1}^\nu + p_i^\nu p_{i+1}^\mu }{p_i \cdot p_{i+1}}\,.
\label{eq:polsum4g}
\end{equation}
In the standard approach, the 10 projectors that single out  the $10$ form factors $F_j$ would then be obtained by inverting the matrix $M_{ij}$.
If one attempts to do this in $d=4$ space-time dimensions, one finds that  $M_{ij}$ is not full
rank and in particular $rank{(M_{ij})} = 8$. This implies, as expected,
that only $8$ tensors are independent when external states are takes in four dimensions.
  In particular, in
order to cover the physical space defined by the external polarizations,
we consider the first seven tensors $T_1,\ldots,T_7$ and the symmetric
combination $T_8+T_9+T_{10}$.  While one can intuitively understand
this to be a good choice due to symmetry reasons, we also show how
construct the basis in a more
systematic way in the appendix~\ref{sec:ggggTphys}.

Let us then follow section~\ref{sec:general} and define
\begin{align}\label{eq:ggggTbar}
  \overline{T}_i ={}& T_i\,, \qquad i=1,...,7 \nonumber \\
  \overline{T}_8 ={}& T_8 + T_9 + T_{10} \nonumber \\
M^{(8\times8)}_{ij} ={}&  \sum_{pol} \overline{T}_i^\dagger \overline{T}_j,
\end{align}
where we still perform
all polarisation sums in $d$ dimensions to compute $M^{(8\times8)}_{ij} $.
The matrix $M^{(8\times8)}_{ij} $ is now invertible in $d=4$ dimensions.
Keeping still generic dependence on $d$, its inverse can be easily computed to be
\begin{equation}
\left(M^{(8\times8)}\right)^{-1}_{ij} = \frac{1}{3(d-1) (d-3) t^2} \left(X^{(0)}_{ij}+ d\, X^{(1)}_{ij}\right)
\end{equation}
with
{\footnotesize
  \begin{align}
X^{(0)}_{ij}={}&\left(
\begin{array}{cccccccc}
 -\frac{8 \left(s^2-4 s u+u^2\right)}{s^2 u^2} & \frac{2 t}{s u} &
   \frac{4 \left(t^2+5 t u+3 u^2\right)}{s u^2} & -\frac{2 t}{s u} &
   \frac{2 t}{s u} & \frac{4 \left(s^2+3 s u-u^2\right)}{s^2 u} &
   -\frac{2 t}{s u} & \frac{-2 t^2+3 t u+3 u^2}{s u} \\[1ex]
 \frac{2 t}{s u} & -2 & \frac{s+2 u}{u} & -1 & 1 & \frac{2 t+u}{s} & -1 &
   s+u \\[1ex]
 \frac{4 \left(t^2+5 t u+3 u^2\right)}{s u^2} & \frac{s+2 u}{u} &
   -\frac{2 \left(t^2+4 t u+u^2\right)}{u^2} & \frac{t-u}{u} & \frac{s+2
   u}{u} & -\frac{(2 s+u) (s+2 u)}{s u} & \frac{t-u}{u} & \frac{t
   (t-u)}{u} \\[1ex]
 -\frac{2 t}{s u} & -1 & \frac{t-u}{u} & -2 & -1 & \frac{2 s+u}{s} & 1 &
   t \\[1ex]
 \frac{2 t}{s u} & 1 & \frac{s+2 u}{u} & -1 & -2 & \frac{2 t+u}{s} & -1 &
   s+u \\[1ex]
 \frac{4 \left(s^2+3 s u-u^2\right)}{s^2 u} & \frac{2 t+u}{s} & -\frac{(2
   s+u) (s+2 u)}{s u} & \frac{2 s+u}{s} & \frac{2 t+u}{s} & \frac{2
   \left(2 s^2-2 t u-3 u^2\right)}{s^2} & \frac{2 s+u}{s} & -\frac{t (2
   t+u)}{s} \\[1ex]
 -\frac{2 t}{s u} & -1 & \frac{t-u}{u} & 1 & -1 & \frac{2 s+u}{s} & -2 &
   t \\[1ex]
 \frac{-2 t^2+3 t u+3 u^2}{s u} & s+u & \frac{t (t-u)}{u} & t & s+u &
   -\frac{t (2 t+u)}{s} & t & t^2 \nonumber \\[1ex]
\end{array}
\right)\\
X^{(1)}_{ij} ={}& \left(
\begin{array}{cccccccc}
 \frac{3 (d+2)}{t^2}+\frac{12}{s^2}-\frac{12}{s u}+\frac{12}{u^2} & -\frac{3}{s+u} & \frac{3
   s (2 s+u)}{t u^2} & \frac{3}{s+u} & -\frac{3}{s+u} & \frac{3 u (s+2
   u)}{s^2 (s+u)} & \frac{3}{s+u} & 0 \\[1ex]
 -\frac{3}{s+u} & 3 & 0 & 0 & 0 & 0 & 0 & 0 \\[1ex]
 \frac{3 s (2 s+u)}{t u^2} & 0 & \frac{3 s^2}{u^2} & 0 & 0 & 0 & 0 & 0 \\[1ex]
 \frac{3}{s+u} & 0 & 0 & 3 & 0 & 0 & 0 & 0 \\[1ex]
 -\frac{3}{s+u} & 0 & 0 & 0 & 3 & 0 & 0 & 0 \\[1ex]
 \frac{3 u (s+2 u)}{s^2 (s+u)} & 0 & 0 & 0 & 0 & \frac{3 u^2}{s^2} & 0 &
   0 \\[1ex]
 \frac{3}{s+u} & 0 & 0 & 0 & 0 & 0 & 3 & 0 \\[1ex]
 0 & 0 & 0 & 0 & 0 & 0 & 0 & 0 \\
\end{array}
\right)\,. \label{eq:M8x8}
\end{align}
}

This matrix allows us to define $8$ intermediate projectors
$$P^{(8\times8)}_i =  \sum_{j=1}^8 \left( M^{(8\times8)}_{ij}\right)^{-1} \, \overline{T}_j^\dagger,$$
which can now by used to find the last two orthogonal tensors $\overline{T}_9$ and $\overline{T}_{10}$
following the prescription given in eq.~\eqref{eq:orthogonalTj}.
This just amounts to removing their projection along the independent tensors as follows
\begin{equation}
\overline{T}_9 = T_9 - \sum_{i=1}^8 \left( P^{(8\times8)}_i\,  T_9 \right) \overline{T}_i\,,
\qquad
\overline{T}_{10} = T_{10} - \sum_{i=1}^8 \left( P^{(8\times8)}_i\,  T_{10} \right) \overline{T}_i\,. \label{eq:newT9T10}
\end{equation}

 As long as the first 8
tensors $\overline{T}_1,\ldots,\overline{T}_{8}$ cover the full space
spanned by the external polarizations, we can forget about these last two
tensors $\overline{T}_9$ and $\overline{T}_{10}$ for any physical
calculation with four dimensional external helicity states.  We can
easily prove this by showing that $\overline{T}_9$ and
$\overline{T}_{10}$ vanishing when contracted with
external helicity states, combined with the fact that -- by
construction -- the last two tensors are completely orthogonal to the
first eight.  Indeed, if we insist in considering all 10 tensors $\overline{T}_i$
and define the matrix
\begin{equation}
\overline{M}_{ij} =  \sum_{pol} \overline{T}_i^\dagger \overline{T}_j\,, \qquad \mbox{for} \;\; i,j = 1,...,10\,,
\end{equation}
it's inverse can be easily computed to be
\begin{equation}
\left(\overline{M}\right)^{-1}_{ij} =\left(
\begin{array}{ccc}
\left(M^{(8\times8)}\right)^{-1}_{ij} & 0 & 0 \\
0 & \frac{2}{(d-3)(d-4)} & \frac{1}{(d-3)(d-4)} \\
0 & \frac{1}{(d-3)(d-4)} & \frac{2}{(d-3)(d-4)}
\end{array}
\right)\,. \label{eq:matrixproj4g}
\end{equation}
As expected, the matrix is in block-diagonal form, such that the two dependent projectors (and tensors) completely decouple from the
relevant ones.

It is thus straightforward to check that the tensors
$\overline{T}_1,\ldots,\overline{T}_{8}$ form a complete physical
basis, by fixing the helicities of the gluons in all possible ways and
showing that the two additional tensors always evaluate to zero.  In other words we have
\begin{equation}
  \overline{T}_9^{\lambda_1,\lambda_2,\lambda_3,\lambda_4} =
  \overline{T}_{10}^{\lambda_1,\lambda_2,\lambda_3,\lambda_4} = 0\,, \label{eq:T9eqT10eq0}
\end{equation}
for any choice of the external helicities.  As before, we indicate
with $\lambda_i=\pm$ the helicity of the $i$-th gluon.  In particular,
we can use the spinor-helicity formalism and
parametrise the polarisation vector of gluon $i$ as follows
\begin{align}
\epsilon_{i,+}^\mu = \frac{\langle q_i | \gamma^\mu | i ] }{\sqrt{2} \langle q_i i \rangle}\,, \qquad
\epsilon_{i,-}^\mu = \frac{[q_i | \gamma^\mu | i \rangle }{\sqrt{2} \langle i q_i \rangle}\,, \label{eq:polglue}
\end{align}
where $q_i$ is the gauge fixing four-momentum which is chosen
respecting eq.~\eqref{eq:polsum4g}.  Equations~\eqref{eq:T9eqT10eq0}
can thus be shown to be correct for any helicity configuration.  A simple way to do this is by means of the momentum twistor
parametrisation for the spinor products~\cite{Hodges:2009hk,Badger:2016uuq}.  This
proves that the tensors $\overline{T}_9$ and $\overline{T}_{10}$ are
irrelevant and thus $\overline{T}_1,\ldots,\overline{T}_{8}$ form a
complete basis of physical tensors for all helicity amplitudes with
four-dimensional external states.

Before moving to the next example, let us see how we expect the construction to generalise if we
consider massive external bosons instead of massless ones, i.e. for example for the
production of two $Z/W$ bosons in gluon fusion $gg \to ZZ$ or $gg \to W^+ W^-$.
If we limit ourselves to vector-like coupling (which is enough if we are only interested
in massless quark loops up to two loops, see for example the discussion in~\cite{Caola:2015ila,vonManteuffel:2015msa}), there are
$20$ independent tensor structures in $d$ dimensions.
On the other hand, having two gluons and two massive vector bosons, and taking parity
invariance under account, tells us that there should be $(2^2 \times 3^2) / 2 = 18$ independent
tensor structures when only four-dimensional external states are considered. This
matches indeed the $18$ independent combinations of form factors which were 
 identified a posteriori in~\cite{Caola:2015ila,vonManteuffel:2015msa}.
We do not construct the tensors explicitly here, but we want to stress that, more recently,
the same number of independent
structures has been obtained with a different approach in~\cite{Davies:2020lpf}. There,
the $20$ independent tensors have been orthonormalised in $d=4$ using the well-known
Gram-Schmidt procedure, which has allowed the authors to construct $18$
independent tensors in the limit $d \to 4$. 

\section{Example 2: $q\bar{q} \to gg$ scattering}
\label{sec:qqgg} \setcounter{equation}{0}
\numberwithin{equation}{section}

Let us repeat the same exercise for the production of two gluons in $q\bar{q}$ annihilation
$$q(p_1) + \bar{q}(p_2) + g(p_3) + g(p_4) \to 0\,.$$
Since all external states are massless, we use the same Mandelstam invariants defined in eq.~\eqref{eq:mandelstam4}.
We start again from the generic $d$-dimensional decomposition
for the amplitude,
 and impose transversality for the external
gluons $\epsilon_i \cdot p_i =0$ with $i=3,4$.
In order to further simplify the problem, we fix the gauge for the external gluons
such that $\epsilon_3 \cdot p_2 = \epsilon_4 \cdot p_1 = 0$, which implies for the gluons
the polarisation sums
\begin{align}
\sum_{pol} \epsilon_3^\mu \epsilon_3^\nu = - g^{\mu \nu} + \frac{p_3^\mu p_{2}^\nu + p_3^\nu p_{2}^\mu }{p_2 \cdot p_3}\,,
\quad
\sum_{pol} \epsilon_4^\mu \epsilon_4^\nu = - g^{\mu \nu} + \frac{p_4^\mu p_{1}^\nu + p_4^\nu p_{1}^\mu }{p_2 \cdot p_3}\,.
\label{eq:polsumqqgg}
\end{align}

With this choice, we are left with 5 independent tensors structures in $d$ dimensions.
A common choice is~\cite{Glover:2003cm}:
\begin{align}
&T_1 =  \bar{u}(p_2) \slashed{\epsilon}_3 u(p_1) \epsilon_4 \cdot p_2 \,, \qquad
T_2 =  \bar{u}(p_2) \slashed{\epsilon}_4 u(p_1) \epsilon_3 \cdot p_1 \,, \qquad
T_3 = \bar{u}(p_2) \slashed{p}_3 u(p_1) \epsilon_3 \cdot p_1 \, \epsilon_4 \cdot p_2\,, \nonumber \\
&T_4 = \bar{u}(p_2) \slashed{\epsilon}_4 \slashed{p}_3 \slashed{\epsilon}_3 u(p_1)\,, \qquad
T_5 = \bar{u}(p_2) \slashed{\epsilon}_3 \slashed{p}_3 \slashed{\epsilon}_4 u(p_1)\,. \label{eq:tensorsqqgg}
\end{align}
Now if we  count the different helicities for the process we get $2^3 = 8$, which should be divided by $2$ due to parity invariance,
such that we expect to have $4$ helicities and $4$ different tensors to describe them. We then expect that one of the tensors
above should be linearly dependent when four-dimensional external
states are considered.
In order to verify that this is the case,
we define the matrix
$$M_{ij} = \sum_{pol} T_i^\dagger T_j$$
where all polarisation sums are done in $d$-dimensions as in eq.~\eqref{eq:polsumqqgg}
and verify easily that it is indeed not full rank in $d=4$, where
$rank{(M_{ij})} = 4$ instead.
In order to find a subset of tensors that spans the full physical
space it is convenient to trade $T_4$ and $T_5$ for their symmetric combination and take as
four independent tensors
\begin{align}
&\overline{T}_i = T_i \,,\quad  i=1,...,3\,, \qquad \overline{T}_4 = \bar{u}(p_2)  \slashed{p}_3 u(p_1) \eps_3 \cdot \eps_4\,.
\end{align}
The fact that this is indeed the right combination can be justified by a similar argument as the one used to pick
the eighth tensor in Eq.~\eqref{eq:ggggTbar} (see also the discussion in the appendices).

As before, in order to define the fifth tensor $\overline{T}_5$ we start from the matrix
$$M^{4 \times 4}_{ij} = \sum_{pol} \overline{T}^\dagger_i \overline{T}_j\,, \qquad i,j = 1,...,4,$$
whose inverse reads
\begin{equation}
\left( M^{4 \times 4} \right)^{-1}_{ij} = \frac{X_{ij}}{(d-3)(s+u)} \,,
\qquad
X_{ij} =  \left(
\begin{array}{cccc}
 -\frac{u}{2 s^2} & 0 & -\frac{u}{2 s^2 (s+u)} & 0 \\
 0 & -\frac{u}{2 s^2} & \frac{u}{2 s^2 (s+u)} & 0 \\
 -\frac{u}{2 s^2 (s+u)} & \frac{u}{2 s^2 (s+u)} & -\frac{d u^2+4 s^2+4 s u}{2 s^2 u (s+u)^2} & \frac{2 s+u}{2 s u (s+u)} \\
 0 & 0 & \frac{2 s+u}{2 s u (s+u)} & -\frac{1}{2 u} \\
\end{array}
\right)\,.
\end{equation}
We then use this matrix to  define the  four intermediate projectors
$$P^{4 \times 4}_i =  \sum_{j=1}^4 \left( M^{(4\times4)}_{ij} \right)^{-1} \, \overline{T}_j^\dagger\,,$$
such that the fifth orthogonal (irrelevant) tensor can be chosen as
\begin{equation}
\overline{T}_5 = T_5 - \sum_{i=1}^4 \left( P^{4 \times 4}_i T_5 \right) \overline{T}_i\,.
\end{equation}
Once more, this achieves effectively a block decomposition of the projector matrix which now reads
\begin{equation}
\overline{M}_{ij} =  \sum_{pol} \overline{T}_i ^\dagger\overline{T}_j\,, \qquad \mbox{for} \;\; i,j = 1,...,5\,,
\end{equation}
\begin{equation}
\left(\overline{M}\right)^{-1}_{ij} = \frac{1}{(d-3)(s+u)} \left(
\begin{array}{cc}
X_{ij} & 0  \\
0 &  -\frac{1}{2u(d-4)}
\end{array}
\right)\,. \label{eq:matrixprojqqb2g}
\end{equation}

Note that this ensures that we only need to perform 4 contractions and not five in order to obtain the full
information required to reconstruct the helicity amplitudes.
For completeness, we write down the explicit expression for the fifth tensor in terms of the original ones
\begin{align}
\overline{T}_5 &= T_5 -\frac{u}{s}T_1 +\frac{u}{s}T_2 -\frac{2}{s}T_3+T_4\,,
\end{align}
which is identically zero when evaluated for four-dimensional external states, for any combination of helicities.
Recently, these projector operators have been successfully applied for the calculation of the three-loop QCD corrections to the
production of two photons in quark-antiquark annihilation~\cite{Caola:2020dfu}.

\section{Example 3: $q\bar{q} \to Q \bar{Q}$ scattering}
\label{sec:qqQQ} \setcounter{equation}{0}
\numberwithin{equation}{section}

The attentive reader might be rather disappointed at this point. In the previous two examples
we have indeed achieved a simplification,
which nevertheless appears to be mainly of aesthetic nature: going from $10$ to $8$ tensors for four-gluon scattering
or from $5$ to $4$ for $q\bar{q} \to gg$ does not seem to be  impressive.
As we will see in the next sections, more impressive simplifications happen when considering the scattering of
$n \geq 5$ particles. Moreover, for what concerns four-particle scattering,
the fact that the new projectors are smooth as $d \to 4$ and their number matches one-to-one
the number of independent helicity amplitudes, could be already a reason for
satisfaction.
Nevertheless, in order to convince also the most demanding reader that this approach is
worth pursuing and conceptually more appropriate also when dealing with four-particle scattering,
in this section we discuss a four-point scattering amplitude where substantial simplifications can be achieved.

The prototypical example of what we want to show is the  production of a pair of quarks in $q\bar{q}$ scattering
$$q(p_1) + \bar{q}(p_2) + Q(p_3) + \bar{Q}(p_4) \to 0\,.$$
This example is particularly interesting because,
if one insists in working  in CDR,
it is not  possible to find a \emph{finite number} of tensor structures which span the whole space
at every number of loops.
The reason is simply that the algebra of the $\gamma$-matrices in $d$-dimensions is not closed.
Indeed, following the standard approach to compute $q\bar{q}Q\bar{Q}$ scattering up to two loops, the following $6$ tensor
structures would be needed~\cite{Glover:2004si,Ahmed:2019qtg}
\begin{align}
&T_1 =  \bar{u}(p_2) \gamma_{\mu_1} u(p_1) \, \bar{u}(p_4) \gamma^{\mu_1} u(p_3) \,, \nonumber \\
&T_2 =  \bar{u}(p_2) \slashed{p}_3 u(p_1) \, \bar{u}(p_4) \slashed{p}_1 u(p_3) \,, \nonumber \\
&T_3 =  \bar{u}(p_2) \gamma_{\mu_1} \gamma_{\mu_2} \gamma_{\mu_3} u(p_1) \, \bar{u}(p_4) \gamma^{\mu_1} \gamma^{\mu_2} \gamma^{\mu_3} u(p_3) \,, \nonumber \\
&T_4 =  \bar{u}(p_2) \gamma_{\mu_1} \slashed{p}_3 \gamma_{\mu_3}u(p_1) \, \bar{u}(p_4) \gamma^{\mu_1} \slashed{p}_1\gamma^{\mu_3} u(p_3) \,, \nonumber \\
&T_5 =  \bar{u}(p_2) \gamma_{\mu_1}  \gamma_{\mu_2} \gamma_{\mu_3} \gamma_{\mu_4} \gamma_{\mu_5}u(p_1) \,
\bar{u}(p_4) \gamma^{\mu_1} \gamma^{\mu_2} \gamma^{\mu_3}\gamma^{\mu_4} \gamma^{\mu_5}u(p_3) \,, \nonumber \\
&T_6 =  \bar{u}(p_2) \gamma_{\mu_1}  \gamma_{\mu_2} \slashed{p}_3  \gamma_{\mu_4} \gamma_{\mu_5}u(p_1) \,
\bar{u}(p_4) \gamma^{\mu_1} \gamma^{\mu_2} \slashed{p}_1\gamma^{\mu_4} \gamma^{\mu_5}u(p_3) \,,
\end{align}
but more would be needed at a higher number of loops.

It should be clear that all of these structures cannot be independent
when the external states are four-dimensional.  Indeed, as argued more
extensively in Appendix~\ref{sec:spinor-chains-with}, it turns out that one only needs
to consider tensors involving \emph{spinor chains} which are \emph{independent in
four dimensions}.
In particular, the 6 structures presented above  can be all related to the first two
tensors $T_1$ and $T_2$ by use of four-dimensional Fiertz identities, hence we
only need to consider $T_1$ and $T_2$.  Alternatively, we can also
start with the full set of tensors and, using the same procedure
illustrated in the previous sections, show that they can be replaced
with only two physical tensors when external four-dimensional states
are specified.
Indeed, let us define once more the matrix
$$M_{ij} = \sum_{pol} T^\dagger_i T_j\,.$$
It is easy to verify that in $d=4$ the $rank(M_{ij}) = 2$.
We define therefore the two independent tensors
$$\overline{T}_i = T_i\,, \quad i=1,2$$
and the $2 \times 2$ matrix $$M_{ij}^{2 \times 2} = T_i^\dagger T_j\,, \qquad i,j=1,2\,,$$
which now has a smooth inverse in $d=4$

\begin{equation}
\left( M^{2\times2} \right)^{-1}_{ij} =  \frac{1}{d-3} X_{ij}\,, \quad X_{ij} =  \left(
\begin{array}{cc}
 \frac{1}{4 s^2} & \frac{s+2 u}{4 s^2 u (s+u)} \\
 \frac{s+2 u}{4 s^2 u (s+u)} & \frac{d s^2-2 s^2+4 s u+4 u^2}{4 s^2 u^2 (s+u)^2} \\
\end{array}
\right)\,.
\end{equation}
It is worth comparing these expressions with the standard $d$-dimensional projectors derived
in eq.(2.15) of~\cite{Glover:2004si}, to appreciate the gain in simplicity.

As for the previous examples, we use this matrix to define the intermediate projectors
\begin{equation}
P^{2\times2}_i =  \sum_{j=1}^2 \left( M^{(2\times2)}_{ij} \right)^{-1} \, \overline{T}_j^\dagger,
\end{equation}
which allows us to decouple the irrelevant tensors by the projection
$$\overline{T}_i = T_i - \sum_{j=1}^2 \left( P^{2\times2}_j T_i \right) \overline{T}_j\,, \quad \mbox{for} \quad i \geq 3\,.$$
For the computation of the helicity amplitudes, all these extra tensors can be neglected.
We stress once more that, if we insisted in using the $d$-dimensional tensors,
their number would increase with the number
of loops, while with our approach two tensors and two form factors are
sufficient to obtain the helicity amplitudes at any number of
loops.
Finally, we can verify that the irrelevant tensors are all zero when we fix the polarisations of the
external quarks in all possible ways, as expected.


\section{Tensor decomposition for $n$-point amplitudes}
\setcounter{equation}{0}
\numberwithin{equation}{section}\label{sec:npoint}

The upshot of the previous sections can be summarised as follows: in order to exploit the simplifications
that arise from treating external states in four space-time dimensions in the 't~Hooft-Veltman scheme,
one only needs identify how many tensor structures remain independent when four-dimensional external states
are considered. We have seen that this number corresponds to the number of independent helicity amplitudes
for the process studied.
Then, to all practical purposes,
one can simply throw away all the linearly dependent ones, which we referred to as \emph{irrelevant tensor structures}.
We have in fact shown, both in complete generality and with many explicit examples, that these irrelevant tensors
can be decoupled from the relevant ones through a redefinition of the tensor basis, which amounts to choosing
irrelevant tensors which are zero when evaluated for any combination of helicity for the external states.
Therefore, the only non-trivial step in extending this procedure to the scattering of $n \geq 5$ particles consists
in determining a priori which tensors are independent in four dimensions.
In principle, one could of course start with the full set of
$d$-dimensional tensors $T_j$ and study the rank of the corresponding projector matrix $M_{ij}$ in eq.~\eqref{eq:matrixproj}.
While this is doable, it can become very soon impractical due to obvious combinatorics arguments.

Fortunately, it turns out that we do not need to go through the enumeration of the $d$-dimensional tensor structures
at all for $n \geq 5$ scattering. The crucial point is that, starting at five points, the external momenta naturally provide us
with four independent vectors to parametrise the helicity amplitudes.
This, combined with the fact that only spinor chains which are
independent in $d=4$ are needed (see
Appendix~\ref{sec:spinor-chains-with}), allows us to decompose
every Lorentz covariant object that appears in the calculation in terms of the of four independent
four-dimensional external momenta
and determine the independent tensors using four-dimensional algebra
only.
This is not true for $n \leq 4$, where we only have three or fewer
independent momenta and we need to introduce extra tensorial
structures (as $\gamma^\mu$, $g_{\mu \nu}$ etc) in order to be able to span the full four-dimensional space.

The simplest case is the scattering of five or more spin-1 massless particles and was treated in detail in~\cite{Peraro:2019cjj}.
In that case, it is immediate to see that the amplitude must
be a rank-$5$ (or higher) tensor, made of the 4 independent momenta $p_1^\mu,...,p_4^\mu$, which implies the existence of $4^5$ different tensor structures.
Applying transversality plus a gauge choice for each external gluon, allows us to go down to $2^5 = 32$ tensor structures, which equals the number of independent helicities.
In~\cite{Peraro:2019cjj} it was shown explicitly that, by projecting on the physical helicity amplitudes, all extra $d$-dimensional
tensor structures would not contribute and could be neglected.
Note that here, unlike the four-point case, we cannot use parity to reduce the number of
independent helicity amplitudes by  a factor two.
The reason is that, starting at 5 points, the helicity amplitudes depend on the parity odd invariant
${\rm tr}5 = {\rm tr}(\gamma_5 \slashed{p}_1\slashed{p}_2\slashed{p}_3\slashed{p}_4)$.
More explicitly, focussing on five-gluon scattering for definiteness,
each helicity amplitude can  be
 separated into a parity even and a parity-odd part
\begin{equation}
\mathcal{A}_{5g}^{\lambda_1,...,\lambda_5}(p_1,...,p_4) = \mathcal{A}_{+}^{\lambda_1,...,\lambda_5}(p_1,...,p_4)
+ {\rm tr}5\;  \mathcal{A}_{-}^{\lambda_1,...,\lambda_5}(p_1,...,p_4)\,,
\end{equation}
where we indicated with $\mathcal{A}_{5g}^{\lambda_1,...,\lambda_5}(p_1,...,p_4)$ the amplitude for the scattering
of five gluons of momenta $p_j$ and helicities $\lambda_j$.
The fact that these amplitudes transform non-trivially under parity,
even after dividing them by an overall spinor phase, does not allow us to restrict our tensor basis to $16$
tensors only. Clearly, the same type of argument can be generalised for the scattering of five or more particles of any type.
In what follows we will  show how to obtain a minimal number of four-dimensional
tensor structures for different types of massless and massive five-point scattering amplitudes, which match
the number of corresponding helicity amplitudes.

\subsection{Example 1: $q \bar{q} \to ggg$ scattering}
We start with the production of three spin-1 massless bosons from a
spinor pair. For definiteness we identify them
with gluons, but clearly the same considerations would apply for photons. We then consider the process
$$q(p_1) + \bar{q}(p_2) + g(p_3) + g(p_4) + g(p_5) \to 0\,.$$
If we were to enumerate all $d$-dimensional tensor structures for this process, we would start
noticing that, by stripping the amplitude of the gluon polarisation vectors, we are left with a rank-3
Lorentz tensor, with one fermion line for the massless quarks. Each tensor will have the form
\begin{equation}
T \sim \epsilon_{3 \mu_3} \epsilon_{4 \mu_4}\epsilon_{5 \mu_5}\; \bar{u}(p_2) \Gamma^{\mu_3\mu_4\mu_5} u(p_1)\,,
\end{equation}
where in $d$-dimensions the tensor $\Gamma^{\mu_3\mu_4\mu_5}$ can be built using combinations of Dirac
$\gamma$  matrices, $g_{\mu \nu}$
and the external momenta $p_i^\mu$, $i=1,...,4$. Note that the maximum number of $\gamma$ matrices is in general
given by the number of particles that can attach on the fermion line, which in turn depends on the number of loops.
Of course, since we have only a finite number of contractions that we can do with the four independent
external momenta, there is a finite number of such structures that we can build. Still, doing this exercise
is very tedious and, as we have seen, not needed, since we only need
to consider spinor chains
that are independent in $d=4$ space-time dimensions.

Indeed, let us start considering the fermion lines $\bar{u}(p_2) \Gamma^{\mu_3\mu_4\mu_5} u(p_1)$.
Their non-trivial part is given by the insertions of (arbitrary numbers of) $\gamma$ matrices
\begin{align}
\bar{u}(p_2) \gamma^{\mu_1} ... \gamma^{\mu_n} u(p_1)\,.
\end{align}
In order to see which structures are effectively allowed to show up in four dimensions,
the crucial observation is that
we can decompose any $\gamma$ matrix as
\begin{equation}
\gamma^{\mu} = a_1 p_1^\mu + ... + a_4 p_4^\mu\,,
\label{eq:decomposegamma}
\end{equation}
where the coefficients $a_j$ will in general depend on the external invariants
and on all possible combinations of strings of $\slashed{p}_i$. Note that this is a non-trivial point.
In fact, while the external states are assumed to be four-dimensional, the general prescription for 't Hooft-Veltman scheme
implies that the
$\gamma$-matrices in the spinor chain be $d$-dimensional. Nevertheless, one can  show that when four-dimensional external states
are considered, one only needs to consider spinor chains which are
independent when also all $\gamma$ matrices are in four dimensions (see appendix~\ref{sec:spinor-chains-with}
for a more detailed discussion).
This implies that, for the purpose of finding a physical tensor basis, the only independent structures that we need to consider are of the form
\begin{equation}
T \sim \epsilon_{3 \mu_3} \epsilon_{4 \mu_4}\epsilon_{5 \mu_5}\;  \bar{u}(p_2) \slashed{p}_{i_1} ... \slashed{p}_{i_n} u(p_1)  p_{j_3}^{\mu_3} p_{j_4}^{\mu_4} p_{j_4}^{\mu_4} \,.
\end{equation}
Now let us first focus on the fermion line. Clearly the indices $i_1,...,i_n$ can be only $i=3,4$, otherwise we could anti-commute
the momenta to the left or to the right and use Dirac
equation to get rid of them. Also, since both $p_1$ and $p_2$ are massless,
helicity conservation along the quark line implies that
we can only have one occurrence of each,
which limits the possible structures to
\begin{equation}
T \sim \epsilon_{3 \mu_3} \epsilon_{4 \mu_4}\epsilon_{5 \mu_5}\; \bar{u}(p_2)\slashed{p}_{3,4} u(p_1)  p_{j_3}^{\mu_3} p_{j_4}^{\mu_4} p_{j_5}^{\mu_5}
\,.
\end{equation}
A simple counting shows that these are $2 \times 4^3 = 128$ different structures.
Now we can use the physical constraints imposed by gauge invariance and transversality of the external gluons to
show that
many of these structure do not contribute. In particular, for each gluon we can impose
$\epsilon_i \cdot p_i = 0$
and $\epsilon_i \cdot p_j = 0$
leaving  $2 \times 2^3 = 16$ structures.
As expected,
the number matches  the number of  helicity amplitudes for the process ($2$ for the quark-line times $8$ for the three
external gluons).
Once an explicit gauge choice is performed (or Ward identities are imposed), the corresponding $16$ projectors
can be obtained inverting a rather simple $16 \times 16$ matrix, which can be easily done with any
computer algebra system.

It should be straightforward to see how to extend these considerations in the case of massive spin-1
external particles. For simplicity, we assume that we are working in a parity invariant theory as QCD or that,
for other considerations, we expect the amplitude not to have any explicit dependence on parity-odd tensor structures.
For each massive boson, we are allowed to impose only the transversality constrain $\epsilon_i \cdot p_i = 0$,
leaving one more possible choice for the momenta to contract the corresponding polarisation vector.
So, for example, if we are interested in the production of two massive vector bosons and a gluon
in $q \bar{q}$ annihilation $q\bar{q} \to VV g$, we will be left with an upper bound of
$2 \times 2 \times 3^2 = 36$ different tensor structures, corresponding to the $36$  helicity amplitudes.

%

\subsection{Example 2:  $q \bar{q} \to Q \bar{Q} g$ and $q \bar{q} \to Q \bar{Q} V$ scattering}
We conclude showing what our method produces when we add an external massless or massive spin-1
particle to the $q\bar{q}Q\bar{Q}$ scattering considered in section~\ref{sec:qqQQ}.
We remind the reader that this case was particularly interesting since one cannot find a
finite number of tensor structures
that
span the full space if one insists in working in CDR.
We showed instead that, by restricting the problem
to four-dimensional external states, two tensors are enough to characterise the helicity amplitudes at
any number of loops.

Let us start by studying what happens when we add an external massless gluon, i.e.\ we consider the process
$$q(p_1) + \bar{q}(p_2) + Q(p_3) + \bar{Q}(p_4) + g(p_5) \to 0\,.$$
As in the previous section, we start analysing the fermion lines. We repeat the same arguments
 independently for the two fermion lines and we find the only possible combinations in four dimensions
\begin{align}
T \sim \epsilon_{5} \cdot p_i\, \bar{u}(p_2)\slashed{p}_{3,4} u(p_1) \, \bar{u}(p_4)\slashed{p}_{1,2} u(p_3)\,.
\label{eq:tensqqQQg}
\end{align}
Without imposing gauge invariance and transversality, these are $2 \times 2 \times 4 = 16$ different tensors,
which become $2\times2\times2 = 8$ once we impose that the external gluon is on-shell and physical, ie
$\epsilon_5 \cdot p_5 = 0$ plus one
more condition to fix its gauge (or impose that the amplitude fulfils QCD Ward identities).
Notice that again this matches the number of helicity amplitudes.

As last case, we study  what happens when the gluon is substituted by a massive external vector boson,
$$q(p_1) + \bar{q}(p_2) + Q(p_3) + \bar{Q}(p_4) + V(p_5) \to 0\,.$$
Again, we assume that all parity-odd contributions coming from possible axial couplings of the massive vector
bosons can be neglected. We stress here that, while this assumption is made  only for exemplification purposes
and similar arguments can be made also in the presence of axial couplings, it is often the
case in massless QCD that axial contributions can be argued away by summing over degenerate
isospin doublets.
All considerations that lead us to the $16$ tensors in eq.~\eqref{eq:tensqqQQg} apply equally well here.
The difference with the case of a massless gluon is that we can only impose one condition on the massive vector boson,
namely transversality $\epsilon_5 \cdot p_5 = 0$. We are
 therefore left with $2 \times 2 \times 3 = 12$ structures, which of course match the number of helicity amplitudes for the problem.

Finally, for both the massless and massive vector boson case, the explicit projectors can be derived
by inverting very simple $8 \times 8$ or $12 \times 12$ matrices, respectively.
The inversion can be obtained with any standard
algebra system. As for the previous cases, we do not present an explicit choice here since,
 depending on the process analysed, different
 choices of gauge can be more or less convenient.

\section{Conclusions}
\label{sec:conclusions} \setcounter{equation}{0}
\numberwithin{equation}{section}
In this paper, we  proposed a new tensor decomposition for
bosonic and femionic scattering amplitudes, that allows us to combine naturally
 the form factor method usually defined in Conventional Dimensional Regularisation,
 with the calculation of helicity amplitudes
 in 't~Hooft-Veltman scheme.
 We have argued that, for an arbitrary number of external legs, one expects a natural
 correspondence between the number of independent helicity amplitudes in the process and the number
 of tensor structures that remain independent when four-dimensional external states
 are considered. We have called these tensors \emph{relevant tensors} and we have described a procedure
 that allows us to effectively decouple them from all remaining $d$-dimensional \emph{irrelevant tensors}.
 While the latter would be required to obtain the value of the scattering amplitudes in CDR, we showed that
 they can be entirely neglected when computing the helicity amplitudes.
 A crucial property is that the \emph{relevant tensors} must be chosen to span the entire vector space where the
 helicity amplitudes are defined.
 After describing the general idea, we have applied it to a large number of explicit examples for both
 bosonic and fermionic amplitudes with $4$ and $5$ external legs.
The new projector operators are smooth in the limit $d \to 4$, and they are substantially simpler than the
ones obtained using the standard approach in CDR.
Importantly, the increase of the number of projectors with the number of legs is
much slower than in CDR, being in particular bounded from above by the number of independent helicity
amplitudes in the problem. Finally, the new projector operators
can be applied on the scattering amplitudes in the very same way as the usual CDR projectors.
We believe that the method proposed here constitutes an improvement compared to the commonly
used approach and
that it could prove useful,  in the future, to compute complicated $2 \to 3$ scattering
amplitudes, as well as lower-multiplicity amplitudes at higher loops.

\section*{Acknowledgements}
We thank F. Caola for many clarifying discussions at different stages of the project.
The work of LT is supported by the Royal Society through Grant URF/R1/191125.

\appendix

\section{Finding a physical basis of tensors for $gg \to gg$}
\label{sec:ggggTphys}

In this paper we have built a basis of physical tensors for several
processes, that are complemented by a set of irrelevant tensor which
can be ignored in the calculation of the helicity amplitudes.  The
physical tensors must, in particular, be a subset of the original
tensors, or linear combinations thereof, that cover the full physical
space spanned by the external polarizations. 
As explained in section~\ref{sec:npoint}, for the scattering of $n \geq 5$ external particles it is always straightforward
to find such a basis by decomposing the tensor structures in terms of 
four independent external momenta. For $n = 4$, instead, this is not always obvious.
Therefore, we show here a
more systematic approach based on the decomposition of momenta in the
parallel and orthogonal space, that has already been successfully used
in other contexts (see e.g.\ ref.~\cite{Mastrolia:2016dhn}).  In
particular, we focus on the $gg \to gg$ example of
section~\ref{sec:gggg}.  Similar arguments can be applied to other
four-point processes.

Because in four-point processes we have three linearly independent
momenta, due to momentum conservation, we can split the physical space
into a three-dimensional part spanned by the external momenta and an
orthogonal part.  Therefore we can, in principle, cover the full
physical space by adding a fourth four-dimensional momentum $v_\perp$
which is orthogonal to the external momenta $p_j$, e.g.\
\begin{equation}
  v_\perp^\mu = \epsilon^{\mu \nu \rho \sigma} p_{1 \nu} p_{2 \rho} p_{3 \sigma}.
\end{equation}
Hence, we may define physical tensors and projectors which only depend on scalar
products of the form $\eps_j\cdot p_k$ and $\eps_j\cdot v_\perp$.   One can also easily verify that a combination with an odd number
of momenta $v_\perp$ is not allowed in a tensor, since it would vanish
when contracted with the amplitude (this is also true in the presence
of external fermions).  A choice of 8 independent physical tensors $\tilde
T_j$ consistent with
the gauge choices of section~\ref{sec:gggg} can be the following
\begin{align}
&\tilde T_1 = \epsilon_1 \cdot p_3\, \epsilon_2 \cdot p_1\, \epsilon_3 \cdot p_1\, \epsilon_4 \cdot p_2 \,, \nonumber \\
&\tilde T_2 = \epsilon_1 \cdot p_3\, \epsilon_2 \cdot p_1\, \epsilon_3 \cdot v_\perp \, \epsilon_4 \cdot v_\perp \,, \quad
\tilde T_3 = \epsilon_1 \cdot p_3\, \epsilon_3 \cdot p_1\, \epsilon_2 \cdot v_\perp \, \epsilon_4 \cdot v_\perp \,,  \nonumber \\
  & \tilde T_4 = \epsilon_1 \cdot p_3\, \epsilon_4 \cdot p_2\, \epsilon_2 \cdot v_\perp \, \epsilon_3 \cdot v_\perp \,, \quad
 \tilde T_5 = \epsilon_2 \cdot p_1\, \epsilon_3 \cdot p_1\, \epsilon_1 \cdot v_\perp \, \epsilon_4 \cdot v_\perp \,, \nonumber \\
& \tilde T_6 = \epsilon_2 \cdot p_1\, \epsilon_4 \cdot p_2\, \epsilon_1 \cdot v_\perp \, \epsilon_3 \cdot v_\perp \,, \quad
\tilde T_7 = \epsilon_3 \cdot p_1\, \epsilon_4 \cdot p_2\, \epsilon_1 \cdot v_\perp \, \epsilon_2 \cdot v_\perp \,, \nonumber \\
&\tilde T_8 = \epsilon_1 \cdot v_\perp \, \epsilon_2 \cdot v_\perp\,  \epsilon_3 \cdot v_\perp \, \epsilon_4 \cdot v_\perp \label{eq:tildetensors4gddim}
\end{align}
In practice, however, the presence of $v_\perp$ would make the
contractions with the amplitude unnecessarily involved.  It is
instead convenient to replace the terms involving orthogonal momentum
$v_\perp$ with terms involving the metric tensor $g^{\mu \nu}$, so
that we can replace the tensors $\tilde T_j$ with linear combinations
of the original tensors $T_j$ defined in Eq.~\eqref{eq:tensors4gddim}.
We can do this by splitting the $d$-dimensional metric tensor
$g^{\mu \nu}$ into a three-dimensional part $g_{[3]}^{\mu \nu}$, which
is its projection into the subspace spanned by the external momenta,
and an orthogonal $(d-3)$-dimensional part $g_\perp^{\mu \nu}$
\begin{equation}
  g^{\mu \nu} = g^{\mu \nu}_{[3]} + g^{\mu \nu}_\perp.
\end{equation}
We can thus exploit the following identities
\begin{align}
  v_\perp^\mu\, v_\perp^\nu \sim{}& g_\perp^{\mu \nu}  \nonumber \\
  v_\perp^{\mu_1}\, v_\perp^{\mu_2}\, v_\perp^{\mu_3}\, v_\perp^{\mu_4} \sim{}& g_\perp^{\mu_1 \mu_2}\, g_\perp^{\mu_3 \mu_4} + g_\perp^{\mu_1 \mu_3} g_\perp^{\mu_2 \mu_4} + g_\perp^{\mu_1 \mu_4}\, g_\perp^{\mu_2 \mu_3}\label{eq:vperpgperp}
\end{align}
where the symbol $\sim$ indicates that the l.h.s.\ and the r.h.s.\
become proportional when contracted with tensors depending on the
$d$-dimensional metric $g^{\mu \nu}$ and the external momenta
$p_j^\mu$ only.  We note that the identities can be easily found
(including the proportionality coefficient, which is irrelevant for
the purposes of this paper) using standard tensor decomposition
methods, although they must be valid also for symmetry reasons.  The
identities above allow us to replace tensors containing $v_\perp$,
with suitable combinations of $g^{\mu \nu}_\perp$.  Moreover, because
$g^{\mu \nu}$ and $g_\perp^{\mu \nu}$ only differ by terms
proportional to the external momenta, which are already accounted for
in our basis of tensors, we are also allowed to replace
$g^{\mu \nu}_\perp$ with $g^{\mu \nu}$.  Putting everything together,
this implies that, starting from the physical basis of tensors
$\tilde T_j$ containing scalar products of the form $\eps_j\cdot v$,
we can make the replacement
\begin{equation}
  \eps_j\cdot v_\perp \, \eps_k\cdot v_\perp \to \eps_j\cdot \eps_k
\end{equation}
in any tensor containing only two scalar products involving $v_\perp$,
that justifies replacing $\tilde T_j \to T_j$ for $j=1,\ldots,7$.  For
tensors with four scalar products involving $v_\perp$ we can similarly
use the second Eq.~\eqref{eq:vperpgperp} to replace
\begin{equation}
  \tilde T_8 \to T_8 + T_9 + T_{10},
\end{equation}
which then yields the same physical tensors defined in
Eq.~\eqref{eq:ggggTbar}.

The same method can be applied to any process with $n\leq 4$ external
legs. One considers a set of $5-n$ orthogonal vectors
$v_{i,\perp}^\mu$ that span the $(5-n)$-dimensional subspace of the
physical four dimensions which is orthogonal to the external
legs.  Thus a basis of tensors is built assuming that they can be
contracted only with external momenta $p_i^\mu$ or the vectors
$v_{i,\perp}^\mu$. This is also true for the $\gamma$ matrices appearing inside
spinor chains, for
the reasons illustrated in Appendix~\ref{sec:spinor-chains-with}.  We note, again, that only combinations with an even
number of orthogonal vectors need to be considered, since odd combinations would
always give zero when contracted with a physical amplitude. Each combination
of the form
\begin{equation}
  v_{i_1,\perp}^{\mu_1}\cdots v_{i_k,\perp}^{\mu_k}
\end{equation}
can be rewritten in terms of the metric tensor $g_\perp^{\mu\nu}$,
which is the restriction of $g_\perp^{\mu\nu}$ in the orthogonal space
and, finally, we can replace $g_\perp^{\mu\nu} \to g^{\mu \nu}$ as
explained above.

We also recall, form the discussion in section~\ref{sec:npoint}, that
these arguments are not needed with $n\geq 5$ external legs, since in
those cases we can always use a subset of four independent external
momenta to span the full physical space.

\section{Spinor chains with four-dimensional external states}
\label{sec:spinor-chains-with}

When dealing with external fermions, we need to build a basis of
tensors involving spinor chains.  In this appendix we show that, as
stated e.g.\ in section~\ref{sec:qqQQ}, when dealing with four-dimensional
external states, we can limit ourself to consider spinor chains that
are independent when restricted to four dimensions.  This is true
despite the fact that the $\gamma$ matrices inside the spinor chain
are $d$-dimensional objects, having a $d$-dimensional Lorentz index,
with $d=4-2\epsilon$.

Because, as we already mentioned, the algebra of the $\gamma$-matrices
in $d$-dimensions is not closed~\cite{Glover:2004si}, in principle
there is no limit to the allowed length of the spinor chains and an
infinite basis of tensors must therefore be considered.  This is, in
particular, relevant when two or more external fermion pairs are
present, because we can contract the Lorentz indices of gamma matrices
belonging to two different spinor chains, which can therefore be
arbitrarily long.  In practice, at each loop order, one can always
limit the length of the spinor chain to the maximum one that can
appear in the diagrams of the process at that perturbative order, but
this still yields a basis of tensors which is much larger than needed,
when dealing with four-dimensional external states.  A different
solution to this problem, that studies how extra-dimensional spinor
indexes decouple from the four-dimensional ones, has also been
proposed in ref.~\cite{Abreu:2018jgq}.

Let us consider a spinor chain of length $n$, of the form
\begin{equation}
  \label{eq:spchain}
  \bar u (p_1)\, \gamma^{\mu_1}\, \gamma^{\mu_2} \cdots \gamma^{\mu_n}\, u(p_2),
\end{equation}
although replacing a spinor $u$ with an anti-spinor $v$ doesn't change
our conclusions.  Here we consider the spinors in
eq.~\eqref{eq:spchain} to be four-dimensional, but the $\gamma$
matrices to be instead $d$-dimensional objects, with a $d$-dimensional
Lorentz index, satisfying the usual anti-commutation relations
\begin{equation}
  \label{eq:gammacomm}
  \left\{ \gamma^{\mu}, \gamma^{\nu} \right\} = 2\, g^{\mu \nu}\, \mathbb{1}
\end{equation}
where $g^{\mu \nu}$ is the $d$-dimensional metric tensor.  In
particular, the Lorentz indexes in eq.~\eqref{eq:spchain} can be
assumed to be contracted with either external momenta and polarization
vectors, or with Lorentz indexes belonging to other spinor chains.  We
can thus split, for each $\gamma$ matrix, contributions from
four-dimensional and $(d-4)$-dimensional indexes with
\begin{equation}
\gamma^\mu = \gamma_{[4]}^\mu + \gamma_{[-2 \epsilon]}^{\mu}
\end{equation}
where $\gamma_{[4]}^\mu$ ($\gamma_{[-2 \epsilon]}^{\mu}$) is an object
that is equal to $\gamma^\mu$ if $\mu$ is a four-dimensional
($-2\epsilon$-dimensional) index and zero otherwise.  We obviously
have
\begin{equation}
  \left\{ \gamma_{[4]}^{\mu}, \gamma_{[-2\epsilon]}^{\nu} \right\} = 0,
\end{equation}
that allows us to anti-commute all the $\gamma_{[-2 \epsilon]}^{\mu}$
matrices to one side of the spinor chain.  This way, up to a
re-labeling of the indexes, the chain in eq.~\eqref{eq:spchain}
becomes a linear combination of contributions of the form
\begin{equation}
  \label{eq:spchainsplit}
  \bar u (p_1)\, \gamma_{[-2 \epsilon]}^{\mu_1} \cdots \gamma_{[-2 \epsilon]}^{\mu_{n_1}} \, \gamma_{[4]}^{\nu_1} \cdots \gamma_{[4]}^{\nu_{n_2}}\, u(p_2).
\end{equation}
with $n_1+n_2=n$.  We can now perform a tensor decomposition of
eq.~\eqref{eq:spchainsplit} with respect to its free (Lorentz)
indexes.  In particular, we can do so in terms of a basis of
four-dimensional vectors $\{e_j^\mu\}_{j=1}^4$ and the
$(-2\epsilon)$-metric tensor $g_{[-2 \epsilon]}^{\mu \nu}$.  Because
the four-dimensional and $(-2 \epsilon)$-dimensional parts are now
completely decoupled from each other, we can also limit ourself to
perform a tensor decomposition of the $\gamma_{[-2 \epsilon]}^{\mu}$
parts alone, in terms of $g_{[-2 \epsilon]}^{\mu \nu}$ only.  By doing
that, we obtain
\begin{equation}
    \bar u (p_1)\, \gamma_{[-2 \epsilon]}^{\mu_1} \cdots \gamma_{[-2 \epsilon]}^{\mu_{n_1}} \, \gamma_{[4]}^{\nu_1} \cdots \gamma_{[4]}^{\nu_{n_2}}\, u(p_2) =  c^{\mu_1\cdots \mu_{n_1}}\, \bar u (p_1)\, \gamma_{[4]}^{\nu_1} \cdots \gamma_{[4]}^{\nu_{n_2}}\, u(p_2)
\end{equation}
where $c^{\mu_1\cdots \mu_{n_1}}$ only depends on
$g_{[-2 \epsilon]}^{\mu \nu}$ and $\epsilon$.  The coefficients
$c^{\mu_1\cdots \mu_{n_1}}$ obviously vanish when contracted with
four-dimensional vectors but they can give a non-vanishing $d$-dependent
contribution when contracted with metric tensors, coming e.g.\ from similar
coefficients in other spinor chains.

We have thus shown that any $d$-dimensional spinor chain of length
$n$, when the external states are four-dimensional, can be re-written in
terms of purely four-dimensional spinor chains (where also the $\gamma$
matrices are in four dimensions) with length $n_1\leq n$.  It is
therefore easy to show (e.g.\ by induction in the length $n$) that
spinor chains in $d$-dimensions, when contracted with four-dimensional
external spinors, are linearly independent \emph{if and only if} they are
independent in four dimensions.  When building a physical basis of
tensors, we are therefore allowed to consider only spinor chains that
are linearly independent when restricted to $d=4$.

\bibliographystyle{bibliostyle}
\bibliography{Biblio}
\end{document}